\documentclass[aps,pre,twocolumn,groupedaddress]{revtex4-1}

\usepackage[DIV12]{typearea}
\usepackage{amsfonts}
\usepackage{amsmath}
\usepackage{graphicx,pgf}
\usepackage{natbib}
\usepackage{enumitem}

\begin{document}


\title{Bounds on the attractor dimension for magnetohydrodynamic channel flow with parallel magnetic field at low magnetic Reynolds number}


\author{R Low}
\email[]{mtx014@coventry.ac.uk}
\author{A Poth\'erat}
\email[]{aa4111@coventry.ac.uk}
\affiliation{Applied Mathematics Research Centre\\Coventry University}


\date{\today}

\begin{abstract}
We investigate aspects of low-magnetic-Reynolds-number flow between two parallel, perfectly insulating walls, in the presence of an imposed magnetic field parallel to the bounding walls. We find a functional basis to describe the flow, well adapted to the problem of finding the attractor dimension, and which is also used in subsequent direct numerical simulation of these flows. For given Reynolds and Hartmann numbers, we obtain an upper bound for the dimension of the attractor by means of known bounds on the nonlinear inertial term and this functional basis for the flow. Three distinct flow regimes emerge: a quasi-isotropic 3D flow, a non-isotropic three-dimensional (3D) flow, and a 2D flow. We find the transition curves between these regimes in the space parameterized by Hartmann number Ha and attractor dimension $d_\text{att}$. We find how the attractor dimension scales as a function of Reynolds and Hartmann numbers (Re and Ha) in each regime. We also investigate the thickness of the boundary layer along the bounding wall, and find that in all regimes this scales as 1/Re, independently of the value of Ha, unlike Hartmann boundary layers found when the field is normal to the channel. The structure of the set of least dissipative modes is indeed quite different between these two cases but the properties of turbulence far from the walls (smallest scales and number of degrees of freedom) are found to be very similar.
\end{abstract}

\pacs{47.65.-5,47.10.Fy,47.27.E,47.10.ad}
\keywords{MHD, channel flow, turbulence, low Rm, dynamical systems}


\maketitle

\section{Introduction}
This paper  focuses on flows of electrically conducting fluids in channels pervaded by a spanwise (i.e. parallel to the bounding walls, rather than perpendicular as in the more familiar case of Hartmann flow) magnetic field with a double aim: 1) to determine the properties of the associated dynamical system, and in particular, an upper bound for the dimension of its attractor, and
2) to derive a functional basis that tightly encompasses the attractor of the system, for subsequent use in highly efficient spectral direct numerical simulations (DNS). Both aims are achieved by deriving the set of least dissipative eigenmodes of the dissipative part of the governing equations.
Before setting out on this task, we familiarize the reader with the key role played by this slightly unusual functional basis in this particular problem and a number of potential others.\\ 
The physical problem is one of the generic configurations where liquid metals flow in devices pervaded by a strong externally imposed magnetic field. It concerns a number of engineering applications in the field of metallurgy and in the nuclear industry, where liquid metals flows are controlled and diagnosed with such fields, or are used to extract heat from nuclear fusion or fission reactors 
\cite{vecha13_pf}.
In these engineering problems and small scale laboratory experiments, the flow falls within the low magnetic Reynolds number (Rm) approximation,  
where the externally imposed magnetic field is considered constant \cite{roberts67}. Its main effect is then to induce electric eddy currents and a Lorentz 
force that acts to eliminate variations of velocity along the magnetic field lines. 
This process has been extensively studied in \cite{moffatt67, dav97, zikanov98_jfm, thess07_jfm, kp10_prl, p12_mhd, pk14_jfm}.
It manifest itself through the presence of very fine boundary layers (Hartmann layers) and highly anisotropic structures. The numerical resolution of the 
boundary layers 
incurs prohibitive computational costs when the magnetic fields becomes high. 
Furthermore, because of the strong anisotropy and Joule dissipation that characterises these flows, Kolmogorov laws for the smallest scales are no longer valid and must be replaced by different scalings for the smallest scales along and across the magnetic field \cite{alemany79}.
On the other hand, at low Rm,  stronger fields incur higher dissipation: this reduces the number of degrees of freedom in the system and therefore, potentially, the computational cost of resolving these flows completely \cite{pa03,pa06_pf}. Recently, a new type of spectral numerical method taking advantage of this property, \cite{dyp09_tcfd,kop15_jcp}, was developed. The number of degrees of freedom is estimated from an upper bound for the dimension of the attractor for the dynamical system associated to the governing equation \cite{doering95}. This more efficient spectral method was constructed in such a way that the flow is represented with a functional basis that encompasses the attractor significantly more tightly than classical bases such as Fourier or Tchebychev bases. 
Because in low-Rm magnetohydrodynamics (MHD), the Lorentz force is exclusively dissipative and linear, such a basis can be found by seeking the eigenmodes of the operator arising from the 
dissipative part of the governing equations, with the boundary conditions of the considered problem \cite{pa03,pa06_pf}. 
In periodic domains and in channels with a transverse magnetic field, 
the derivation of this basis 
provided an upper bound for the attractor dimension, scalings for the smallest 
scales and the thickness of wall boundary layers that could be verified heuristically and numerically. Most importantly, it 
made it possible to calculate turbulent MHD flows in almost arbitrarily high magnetic fields at a moderate computational cost \cite{pdy10_jfm, kop15_jcp}.\\
The problem of channel flows with a uniform spanwise magnetic field has received much recent attention \cite{krasnov08,krasnov08_pf} but no such basis is yet known for it. Consequently, 
upper bounds for the attractor dimension, scalings for the smallest scales and 
the thickness of the boundary layers along the channel walls and are not available. Nor is it possible to perform efficient spectral DNS at high magnetic field.

We therefore set out to answer these question for this geometry by deriving 
the basis of least dissipative modes and analysing its properties.
We first derive analytically the least dissipative eigenmodes (Sec. \ref{sec:eigenmodes}), then numerically calculate their associated eigenvalues (Sec. \ref{sec:numerics}). From these, we deduce an upper bound for the attractor dimension of the system and distinguish three possible regimes: weakly 3D, strongly 3D  and 2D (Sec. \ref{sec:attractor}). Finally from the set of least dissipative modes, we shall extract scalings for the thickness of the boundary layer that develops along the channel walls and for the size of the smallest scale present in the flow (Sec. \ref{sec:scalings}).

\section{Governing equations and procedure for obtaining bounds on the attractor dimension \label{sec:eigenmodes}}

\subsection{Governing equations}
To evaluate the attractor dimension, let us consider the time evolution of the 
flow as given by a dynamical system whose phase space is the space of all 
solenoidal vector-valued functions on the fluid containing region.  This time 
evolution is specified by the Navier-Stokes equations.  We proceed by 
considering the time evolution of an infinitesimal perturbation to a  flow, $\pmb{U}$, that evolves in the neighbourhood of the attractor ($\pmb{U}$ follows the attractor itself). To obtain an upper bound for the attractor dimension, we note that such a perturbation, which we denote $\pmb{u}$, spans a $n$-dimensional infinitesimal volume, which should asymptotically contract to 0 as soon as $n$ is larger than the embedding dimension of the attractor 
\cite{doering95}.
Denoting by $\pmb{u}$ this perturbation, and ignoring higher-order terms, we 
then find the trace of the linearized Navier-Stokes equations, which determine 
the contraction or expansion rate of this volume. Making use of an estimate 
obtained in earlier work for the part of the trace due to inertia \cite{doering95}, we then obtain an upper bound 
to the attractor dimension of this dynamical system at various Reynolds and Hartmann numbers. Note that since the trace of the operator is independent of the basis, the basis of eigenvectors of $\mathcal D_{Ha}$ need not be orthogonal, and in general, it isn't.\\

The physical problem we consider is that of the flow of a fluid of density $\rho$, conductivity $\sigma$, and kinematic viscosity $\nu$, which is confined
between impermeable perfectly electrically insulating walls at $z=\pm L$, and subject to periodic boundary conditions in the $x$ and $y$ directions at 
$x=\pm\pi L$ and $y=\pm\pi L$ in the presence of an applied magnetic field 
 $\pmb{B}=B\pmb{e}_x$. 
We consider the usual Navier-Stokes equation for MHD within the quasi-static 
MHD approximation, which is valid as long as the induced magnetic field remains 
small compared to the externally imposed one \cite{roberts67}. Taking $L$ to be 
the typical distance, and $U$ the typical velocity, the evolution 
of $\pmb U$ within this approximation is written nondimensionally as
\begin{equation}
\begin{split}
\partial_t \pmb{U} &= -\nabla p - \pmb{U}\cdot\nabla \pmb{U}
+\text{Re}^{-1} \left(\Delta - \text{Ha}^2  \Delta^{-1} \partial_{xx} \right)\pmb{U},\\
\nabla.\pmb{U}&=0,
\end{split}
\end{equation}
and the evolution of $\pmb{u}$ is given by \cite{pa03,pa06_pf}:
\begin{equation}
\begin{split}
\partial_t \pmb{u} &= -\nabla p - \pmb{U}\cdot\nabla \pmb{u} - \pmb{u}\cdot\nabla \pmb{U}
+\text{Re}^{-1} \left(\Delta - \text{Ha}^2  \Delta^{-1} \partial_{xx} \right)\pmb{u},\\
\nabla.\pmb{u}&=0,
\end{split}
\end{equation}
where $p$ is the perturbation to the pressure, and $\Delta$ is the Laplacian operator, whose inverse is well-defined for functions satisfying the boundary conditions of interest here, namely
\begin{eqnarray}
\pmb {u}(x,y,\pm 1) &=& 0,\nonumber \\
\pmb {u}(x,y,z)&=&\pmb {u}(x+2 \pi,y+2 \pi,z). 
\end{eqnarray}
The problem is governed by two non-dimensional parameters, the Hartmann and 
Reynolds numbers $\text{Ha}=\sqrt{\sigma/(\rho\nu)}BL$ and $\text{Re}=UL/\nu$ 
respectively. 
The perturbation to the current, $\pmb{j}$, is given by Ohm's law:
\begin{equation}
\pmb{j} = (-\nabla \phi + \pmb{u} \times \pmb{e}_x).
\label{ohm}
\end{equation}
Taking the curl of this twice,  and using the fact that $\nabla\cdot\pmb{j}=0$
yields
\begin{equation}
\pmb{j} = - \Delta^{-1} \partial_x \nabla \times \pmb{u} = -\Delta^{-1} \partial_x \pmb{\omega},
\label{jinvlap}
\end{equation}
where $\pmb{\omega}$ is the vorticity. Finally, $\pmb{j}$ satisfies the boundary condition
\begin{equation}
j_z(x,y,\pm 1)=0.
\end{equation}
We now consider a space spanned by $n$ mutually orthogonal perturbations, and the behaviour of this space as it evolves. Denoting by $\mathcal{A}$ the linearised evolution operator, and by $P_n$ the projection to this space, it can be shown \cite{constantin85_ams,pa03} that the trace of $\mathcal{A}P_n$ satisfies the inequality
\begin{equation}
\text{Tr}(\mathcal{A}P_n) \leq \text{Tr}\left( \left[ \frac{1}{2}\Delta - \text{Ha}^2\Delta^{-1} \partial_{xx}\right] P_n\right) + \frac{n}{2}\text{Re}^2 \label{trdim}.
\end{equation}
To find the attractor dimension, we find the eigenvalues of the operator $\mathcal{D}_{\textrm{Ha}}=1/2 \Delta - \text{Ha}^2\Delta^{-1} \partial_{xx}$, listed in decreasing order (bearing in mind that the eigenvalues of this operator are real and negative, as expected from a purely dissipative operator). Then the lowest value of $n$ which gives a negative value for the upper bound of the trace of $\mathcal{A}P_n$ provides an upper bound on 
the attractor dimension for a given value of the Reynolds number. Finding the 
eigenmodes and eigenvalues of $\mathcal{D}_{\textrm{Ha}}$ in order to obtain 
these bounds is our next aim. We therefore require the solution of the eigenvalue problem for the operator  
$\mathcal{D}_{\textrm{Ha}}$ in a closed box with periodic boundary conditions 
of period $2\pi$ in the $x$ and $y$ directions, and with impermeable, perfectly 
insulating walls at $z=\pm 1$. 

In terms of the non-dimensional variables, this yields the eigenvalue problem
\begin{equation}
\mathcal{D}_{\text{Ha}} \pmb{u}=\lambda \pmb{u},
\end{equation}
where the symmetry of the operator $\mathcal{D}_{\text{Ha}}$ guarantees that 
the eigenvalues are real.
As the Laplacian operator is invertible for these boundary conditions, we can instead take the Laplacian of both sides and consider the problem
\begin{equation}
(\Delta^2 - 2 \text{Ha}^2 \partial_{xx}) \pmb{u} = 2 \lambda \Delta\pmb{u} \label{ev}
\end{equation}
where $\pmb{u}$ satisfies the incompressibility condition $\nabla\cdot\pmb{u}=0$. 
From the periodicity in $x,y$, 
we can consider $\pmb{u}$ to be a sum of terms of the form
\begin{equation}
\pmb{u}=e^{i\pmb{k.x}} \sum_{i \in \{x,y,z\}} Z_i(z)\pmb{e}_i, 
\end{equation}
where $\pmb{k}=k_x\pmb{e}_x+k_y\pmb{e}_y$, $(k_x,k_y) \in \mathbb{Z}^2$, and $\pmb{x}=x\pmb{e}_x+y\pmb{e}_y$. Now consider  a single component of $\pmb{u}$, denoted by $e^{i\pmb{k.x}}Z_i(z)$. For this to be a solution to the eigenvalue problem (\ref{ev}), we must have 
\begin{equation}
Z_i''''-2(\lambda + k^2)Z_i''+(k^4+2\lambda k^2+2\text{Ha}^2k_x^2)Z_i=0,
\end{equation}
where a prime denotes differentiation with respect to $z$.
We seek a solution of the form $Z_i(z)=e^{Kz}$, resulting in the auxiliary quartic equation
\begin{equation}
K^4-2(\lambda + k^2)K^2+(k^4+2\lambda k^2+2\text{Ha}^2k_x^2)=0.
\label{eq:charact}
\end{equation}
Solving this quadratic equation in $K^2$ yields the two roots
\begin{equation}
\begin{split}
K_1^2=\lambda+k^2+\sqrt{\lambda^2-2\text{Ha}^2k_x^2},\\
K_2^2=\lambda+k^2-\sqrt{\lambda^2-2\text{Ha}^2k_x^2},
\end{split}
\end{equation}
and eliminating $\lambda$ from these gives the relation
\begin{equation}
K_1^2K_2^2=k^2(K_1^2+K_2^2)-k^4+2\text{Ha}^2k_x^2.
\label{disp}
\end{equation}
This relation gives one constraint on the allowed roots of the auxiliary equation: other constraints are provided by the boundary conditions on the flow. Once these constraints have been solved to give $K_1$ and $K_2$, we obtain the corresponding eigenvalue from (\ref{eq:charact}):
\begin{equation}
\lambda=\frac{1}{2}(K_1^2+K_2^2)-k^2.
\label{eig}
\end{equation}

From the impermeability and non-slip conditions at $z=\pm 1$, together with incompressibility, we have
\begin{equation}
Z_i(\pm 1)=0 = Z_z'(\pm 1), \label{fbc}
\end{equation}
and
\begin{equation}
ik_x Z_x(z)+ik_y Z_y(z) + Z'_z(z) = 0. \label{inc}
\end{equation}
In addition, we obtain electrical boundary conditions from the current field
$\pmb{j}$,  which is determined by $\Delta \pmb{j}=-\partial_x \pmb{\omega}$ (\ref{jinvlap}). Taking the curl of (\ref{ev}) and considering the $z$-component of $\pmb{j}$ finally gives the boundary condition
\begin{equation}
k_y Z''_x(x,y,\pm 1)=k_x Z''_y(x,y,\pm 1), \label{ebc}
\end{equation}
since we must have $j_z(x,y,\pm 1)=0$.

Modes can conveniently be divided into two classes: those for which $Z_z(z)$ is not identically zero, and so the boundary conditions (\ref{fbc}) must be underdetermined, and those for which $Z_z(z)$ is identically zero, so that non-zero $Z_x(z)$ and $Z_y(z)$ must satisfy the electric boundary condition (\ref{ebc}). By analogy with linear stability theory in hydrodynamics, we call these the  Orr-Sommerfeld (OS) and Squire  modes, respectively. $\lambda$ is the 
exponential decay rate of the corresponding eigenmode under the sole effect of 
dissipation (viscous and Joule). Such a decay would, however, only be observed
 on individual modes and in the absence of inertia. The evolution of more 
complex linear flows can still be expressed as a combination of exponential 
decays \cite{dyp09_tcfd}.
\subsection{Expressions of modes and eigenvalues}
We can now solve the eigenvalue problem to find the modes and corresponding 
eigenvalues explicitly. In addition to providing an upper bound on the 
attractor dimension, the basis formed with these modes can be used to carry out 
numerical simulations of the flows under consideration by means of spectral 
methods so they constitute an important result of their own \cite{kop15_jcp}. 
A laborious calculation shows that the only significant possibilities are 
\begin{enumerate}
\item OS-type modes where one  of $K_1^2$ and $K_2^2$ is positive, and the other 
negative, and where $k_x$ and $k_y$ are not both zero,
\item OS-type modes where both $K_1^2$ and $K_2^2$ are negative, and $k_x$ and $k_y$ are not both zero ,
\item Squire type modes with $k_x=k_y=0$.
\end{enumerate}
In each case above, $|K_1|$ is different from $|K_2|$. Other possible cases are those in which $|K_1|=|K_2|$,
or $K_1^2$ and $K_2^2$ are complex.
 In each of these cases, either 
there is no non-trivial mode at all, or for any given choice of $k_x$ and $k_y$  
there is a mode for just one precisely tuned value of $\text{Ha}$. These 
singular cases are of lesser importance for our purpose but are an interesting 
property of this problem, which is absent when the magnetic field is 
perpendicular to the walls for instance \cite{pa06_pf}. They are briefly described 
in appendix \ref{sec:resonant}. We now restrict our attention to the 
generic case where for a chosen Hartmann number there is a set of solutions to 
the constraints.

In case (1), we denote the roots of the auxiliary quartic by $K_1=\pm1/\delta$ and $K_2=\pm i \kappa_z$, where $1/\delta \neq \kappa_z$. This reflects that real roots correspond to the exponential profile of a boundary layer of thickness $\delta$ near the walls, whereas imaginary ones induce spatial oscillations of wavelength in the bulk of the flow. In this case, in order to have a non-trivial $Z_z$ mode, we require 
\begin{eqnarray}
1/\delta \tanh 1/\delta &=& -\kappa_z \tan \kappa_z \qquad \text{or} \label{eq:solvab_osa}\\
1/\delta \tan\kappa_z &=& \kappa_z \tanh 1/\delta \label{eq:solvab_osb}
\end{eqnarray}
and
\begin{equation}
-\frac{\kappa_z^2}{\delta^2}=k^2(1/\delta^2-\kappa_z^2)-k^4+2\text{Ha}^2k_x^2.
\label{eq:disp_orr}
\end{equation}
If neither $k_x$ nor $k_y$ is zero, then the two possibilities are
\begin{equation}
\begin{split}
Z_z(z)&=-\cos\kappa_z \cosh(z/\delta)+\cosh(1/\delta)\cos(\kappa_z z)\\
Z_x(z)&=i\frac{k_x ( \kappa_z^3 \cosh(1/\delta) \sin \kappa_z-1/\delta^3 \cos \kappa_z \sinh(1/\delta)) }{(1/\delta^2+\kappa_z^2)(k_x^2+k_y^2)}\\
&\times \left(\frac{\sinh(z/\delta)}{\sinh(1/\delta)}- \frac{\sin(\kappa_z z)}{\sin \kappa_z}\right)\\
Z_y(z)&=\frac{1}{k_y}(iZ'_z(z)-k_xZ_x(z))
\end{split}
\end{equation}
and
\begin{equation}
\begin{split}
Z_z(z)&=-\sin \kappa_z \sinh(\kappa_z z) +\sinh(1/\delta) \sin(\kappa_z z)\\
Z_x(z)&=i\frac{k_x(1/\delta^3 \sin \kappa_z \cosh(1/\delta)  + \kappa_z^3 \sinh (1/\delta) \cos \kappa_z)}{(k_x^2+k_y^2)(1/\delta^2+\kappa_z^2)}\\
&\times \left( \frac{\cos(\kappa_z z)}{\cos \kappa_z} - \frac{\cosh(z/\delta)}{\cosh(1/\delta)} \right)\\
Z_y(z)&=\frac{1}{k_y}(iZ'_z(z)-k_xZ_x(z)).
\end{split}
\end{equation}
If $k_x=0$ then from (\ref{ebc}) we immediately have $Z_x(z)=0$, and from (\ref{inc}) we then obtain
\begin{equation}
Z_y(z)=iZ_z'(z)/k_y
\end{equation}
and similarly if $k_y=0$.

In case (1), for each interval of the form $[n\pi/2,(n+1)\pi/2]$ there is one 
value of $\kappa_z$ and a corresponding value of $1/\delta$ satisfying the 
constraints, as for the analogous modes in the case where the magnetic field 
is perpendicular to the walls.

In this case, (\ref{eig})  gives
\begin{equation}
\lambda=\frac{1}{2}\left(\frac{1}{\delta^2} - \kappa_z^2 \right) - k^2.
\end{equation}

In case (2), we have $K_1=i\tilde{\kappa}_z$, and $K_2=i\kappa_z$, where $\tilde{\kappa}_z \neq \kappa_z$. This time the non-trivial $Z_z$ modes are given by
\begin{eqnarray}
\tilde{\kappa}_z \tan \tilde{\kappa}_z &=& \kappa_z \tan \kappa_z \qquad \text{or} \label{eq:solvab_osi1} \\
\tilde{\kappa}_z \tan \kappa_z &=& \kappa_z\tan \tilde{\kappa}_z.
\label{eq:solvab_osi2}
\end{eqnarray}
This yields:
\begin{equation}
\begin{split}
Z_z(z)&=-\cos \kappa_z \cos(\tilde{\kappa}_z z)+\cos \tilde{\kappa}_z \cos (\kappa_z z)\\
Z_x(z&)=i\frac{k_x(\tilde{\kappa}_z^3 \cos \kappa_z \sin \tilde{\kappa}_z  - \kappa_z^3 \cos \tilde{\kappa}_z \sin \kappa_z )}{(k_x^2+k_y^2)(\tilde{\kappa}_z^2-\kappa_z^2)} \\ 
&\times \left(\frac{\sin(\tilde{\kappa}_z z)}{\sin \tilde{\kappa}_z} - \frac{\sin(\kappa_z z)}{\sin \kappa_z}\right)\\
Z_y(z)&=\frac{1}{k_y}(iZ'_z(z)-k_xZ_x(z))
\end{split}
\end{equation}
or
\begin{equation}
\begin{split}
Z_z(z)&=-\sin(\kappa_z)\sin(\tilde{\kappa}_z z)+\sin(\tilde{\kappa}_z) \sin(\kappa_z z)\\
Z_x(z)&=i\frac{k_x(\kappa_z^3 \sin \tilde{\kappa}_z \cos \kappa_z-\tilde{\kappa}_z^3 \sin \kappa_z \cos \tilde{\kappa}_z)}{(k_x^2+k_y^2)(\tilde{\kappa}_z^2-\kappa_z^2)}\\
&\times \left(\frac{\cos(\tilde{\kappa}_z z)}{\cos \tilde{\kappa}_z}- \frac{\cos(\kappa_z z)}{\cos \kappa_z}\right)\\
Z_y(z)&=\frac{1}{k_y}(iZ'_z(z)-k_xZ_x(z))
\end{split}
\end{equation}
when neither of $k_x$ not $k_y$  are zero, and obtain the $Z_z$ and $Z_y$ modes as before if one of them is zero.

This time (\ref{eig}) gives
\begin{equation}
\lambda = -\frac{1}{2} (\kappa_z^2+\tilde{\kappa_z}^2)-k^2.
\end{equation}
In case (2), however, the roots are not as conveniently located as in the previous case; as $\kappa_z$ increases, they can become arbitrarily close together. We will consider the consequences of this in the next subsection.

Finally, we have the Squire modes (case (3)), which occur only when $k_x=k_y=0$. In this case we have $Z_z(z)=0$, and  if $n$ is a positive integer, $Z_x(z)$ and $Z_y(z)$ are given either by
\begin{equation}
Z_{x,y}(z)=\cos((n+1/2)\pi z), 
\end{equation}
where $\lambda=-\frac{1}{2}(n+1/2)^2 \pi^2$
or by
\begin{equation}
Z_{x,y}(z)\sin(n\pi z), 
\end{equation}
where $\quad \lambda =-\frac{1}{2}n^2\pi^2$.
These functions $Z_x$, $Z_y$ and $Z_z$ then provide a functional basis for consideration of flows, which will be
be applied in DNSs analogous to those in  \cite{dyp09_tcfd,kop15_jcp}. The full expression of the functional basis is given in appendix \ref{sec:basis}.

\section{Numerical method and validation \label{sec:numerics}}

In order to find the eigenvalues of the modes and the corresponding values of $k_x$, $k_y$, $\kappa_z$, and $1/\delta$, a numerical approach was required. The approach taken was to find, for each of an increasing family of values of $\text{Ha}$, all the modes and eigenvalues up to a limiting value. 

In case  (1) (one real and one imaginary root), finding the eigenvalues was straightforward; the roots are located in known intervals so that it is easy to find the root in each interval by means of a bisection method. 

\begin{figure}
\begin{center}
\includegraphics[width=0.5\textwidth]{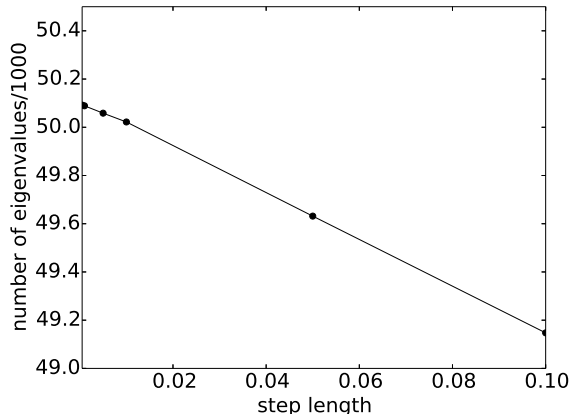}
\caption{Number of eigenvalues found versus step length.}
\label{fig:convergence}
\end{center}
\end{figure}

In the case (2), the roots are not spread out  in such a convenient manner. In 
fact, as the Hartmann number grows and the relevant values of $\kappa_z$  become 
larger, the roots can become arbitrarily close together. It is therefore necessary
 to use a much smaller step length, use Eq. (\ref{disp}) to express  
(\ref{eq:solvab_osi1}) and (\ref{eq:solvab_osi2}) in terms of just one of the roots, 
 and check for a sign change. The number of roots found as the step length is decreased
is shown in Fig \ref{fig:convergence}. We found that to an excellent degree of approximation the
number of roots varied linearly with step length: a linear fit gives
\begin{equation}
n=5.01 \times 10^5 - 9.644 \times 10^4 s
\end{equation}
with  goodness of fit measure $R^2 = 0.9996$,
where $n$ is the number of roots and $s$ is the step length. Extrapolating to $s=0$,
we find that the fraction of roots omitted with $s=0.001$ is about $0.02\%$.
The fact that some roots are omitted means 
that the estimate for $|\text{Tr}(\mathcal{D}_{\textrm{Ha}}P_n)|$ is a slight  
underestimate, but the eigenvalues are very closely spaced, so that although 
some eigenvalues are omitted, the sum of the first $n$ eigenvalues obtained is 
close to the sum of the first $n$ of all eigenvalues.
The numerical methods were implemented in Python, allowing the investigation of a maximum Hartmann number of about $4400$, and numbers of modes in excess of $10^7$. The main obstruction to investigating higher Hartmann number was that of computational time: as a consequence of the short step length required, the calculation of all required modes is very time consuming at high Hartmann number.\\ 
Once obtained, these modes for each Hartmann number considered were then sorted in order of increasing magnitude of $\lambda$, and the modulus of the sum of the first $n$ eigenvalues used as an estimate for $\text{Tr}(\mathcal{D}_{\textrm{Ha}}P_n)$. From (\ref{trdim}), an approximation to the Reynolds number for which $n$ is the dimension of the attractor is then given by $\sqrt{|2\text{Tr}(\mathcal{D}_{\textrm{Ha}}P_n)|/n}$.\\
It is interesting to note that for small values of the Hartmann number, the modes of the first type (one real and one imaginary root) predominate---indeed, there is no contribution from modes of the second type (two imaginary roots) until the Hartmann number exceeds 1.2. But as the Hartmann number grows, the contribution from  modes of the second type grows until at Hartmann numbers exceeding 3000, the modes of the first type are only about a quarter of all those considered.
Such an inhomogeneous distribution of modes was not observed in 3D periodic domains nor in MHD channels perpendicular to $\mathbf B$. A significant consequence is that since the spectral density of the distribution of modes in $(k_x,k_y,\kappa_z)$ space cannot be easily predicted, it is no longer possible to obtain an analytical estimate for the upper bound for $d_{\rm att}$ by means of a simple approximation of the Trace in (\ref{trdim}) by a continuous integral, as in \cite{pa03} and \cite{pa06_pf}. Consequently, it has to be obtained numerically only.

\section{Distribution and physical properties of least dissipative modes \label{sec:attractor}}
\subsection{Spectral distribution of eigenvalues
\label{sec:distrib}}
Even though the least dissipative modes do not give an exact solution of the full system of equations governing the flow evolution, it has been shown that finite combinations of them were able to provide an accurate representation of the actual solution \cite{pdy10_jfm,kop15_jcp}, at least in low-Rm MHD flows. Much can therefore be learned from the flow properties by studying the properties of such finite sets of modes. Since $\lambda<0$, modes can be sorted by growing dissipation rate $\lambda_n$. By construction, the $N$ least dissipative modes are contained within the region delimited by a manifold $\lambda(k_x,\kappa_y,k_z)=\lambda_N$ of the $(k_x,\kappa_y,k_z)$ space. The shape of these manifolds therefore gives a good measure of the flow anisotropy, in particular at small scales. 
From (\ref{eig}), these can be rescaled to a single manifold representing 
surfaces of constant $\lambda/\text{Ha}$ in the $(k_x/\text{Ha},k_y/\text{Ha},\kappa_z/\text{Ha})$ space.\\
Since $k_x=0$ removes the Hartmann number from the situation, the shapes of the 
contours in the $k_x=0$ plane are unaffected by the growing magnetic field.
Consequently, the cross-section of this family of manifolds in planes 
$k_x=$constant is very close to a family of concentric circles, which indicates that  iso-$\lambda$ manifolds are isotropically distributed in in planes perpendicular to the magnetic field direction. This was indeed the case too in channels with a transverse magnetic field and in periodic domains. Discrepancy to anisotropy in these two cases was only due to the discrete distribution of values of wavenumbers perpendicular to $\mathbf B$, which had to be integers. Because 
of the walls at $z=\pm1$, though, $\kappa_z$ spans the solutions of $(\ref{eq:solvab_osa}) ,(\ref{eq:solvab_osb}), (\ref{eq:disp_orr} )$ or  $(\ref{eq:solvab_osi1}) ,(\ref{eq:solvab_osi2}), (\ref{eq:disp_orr})$ rather than the set of integers. This effectively introduces a form of anisotropy in the sense that the sets of wavenumbers in the $x$ and $z$ directions are not identical, but still span the same interval.\\
\begin{figure}[h!]
\begin{center}
\includegraphics[width=0.4\textwidth]{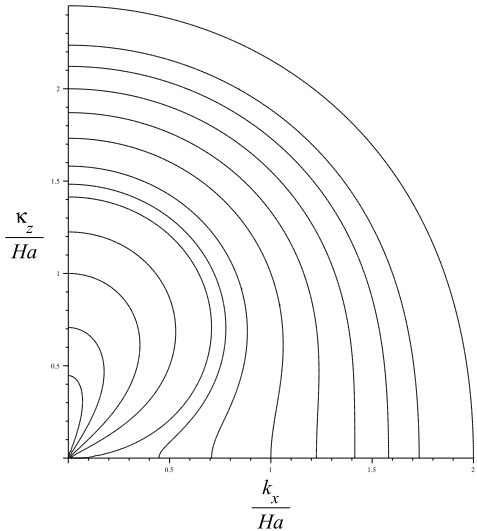}
\caption{Contours of fixed values of $\lambda/$Ha}
\label{isol}
\end{center}
\end{figure}
Cross-sections of the iso-$\lambda$ manifolds in the $k_y=0$ plane tell a 
different story (see Fig. \ref{isol}). 
The geometry of this graph is formally identical to geometries of channel flow 
with transverse magnetic field \cite{pa06_pf}, but with the roles of $\kappa_z$ 
and $k_x$ reversed. Both phenomenologies thus bear strong similarities, but for 
the orientation of the magnetic field. 
For small Ha, the flow is essentially isotropic, $k_x$ and $k_y$ have 
similar behaviours, and the manifolds are spheres.
As Ha increases, the increasing suppression of the $k_x$ modes distorts the 
contours in the $k_y=0$ plane. This effect becomes more pronounced as Ha 
increases, until we obtain situations where the $k_x$ modes are almost entirely 
suppressed.\\
Conversely, for a fixed value of Ha, as the value of the largest eigenvalue (and so the number of modes under consideration) increases, we have the following sequence. Initially, the modes have $k_x$ strongly suppressed, and the flow is essentially 2D. Next, we enter a regime where the curves of constant $\lambda$ pass through the origin; this is the 3D, anisotropic regime. In this regime, all modes are contained outside a cone of axis $\mathbf e_x$, tangent to the manifold at the origin, whose half-angle is easily derived from (\ref{eig}) as $\theta_J=\pi/2-\cos^{-1}(\sqrt{-\lambda}/Ha)$. This phenomenology reflects that in MHD turbulence at high interaction parameter $S=\sigma B^2L/(\rho U)$, all energy-containing modes are expelled from the Joule cone \cite{knaepen08_arfm,p12_mhd}.
Finally, the contours split away from the origin and we reach the regime of weakly anisotropic 3D flow, which becomes more closely isotropic as the contours approach a semi-circular shape. We also see this from a different perspective in the following subsection.\\
\subsection{Upper bound for the attractor dimension \label{sec:bounds}}
\begin{figure}[h!]
\begin{center}
\includegraphics[width=0.5\textwidth]{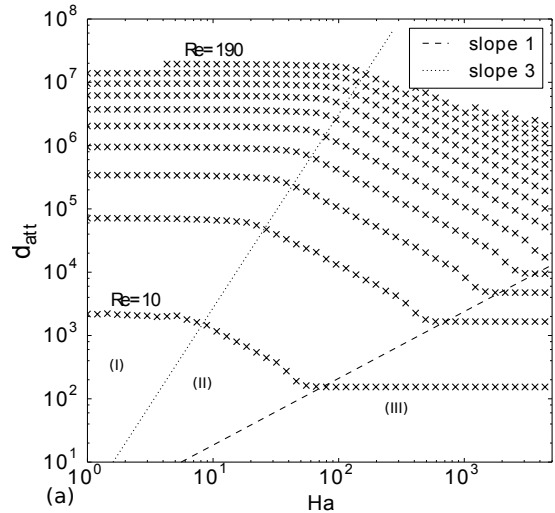}
\includegraphics[width=0.5\textwidth]{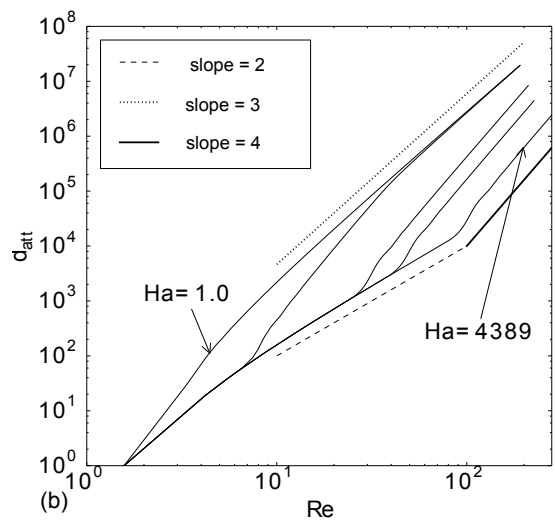}
\caption{Attractor dimension (a) as a function of Ha for Re ranging from 10 (lowest data set) to 270 (highest data set) in steps of 20 and (b)
as a function of Re for Ha=1.0 (leftmost curve), 26.62, 6, 410.2 (from left to right), 4389 (rightmost curve)  (b).}
\label{attdim}
\end{center}
\end{figure}

Now consider the plot of attractor dimension vs Hartmann number for fixed Reynolds number. Figure \ref{attdim} plots the dimension of the attractor for Reynolds numbers starting at $10$ and increasing in steps of $20$, for Hartmann number starting at $1$ and increasing in multiplicative steps of 1.2 up to a maximum value of approximately 4400.
We note that as in \cite{pa03} this plot divides the plane up into three regions: a 3D quasi-isotropic region (I), a 3D anisotropic region (II), and a 2D region (III), corresponding to the classification given above.\\
For a fixed Reynolds number, and increasing Ha, the dimension initially 
depends only weakly on Hartmann number in the quasi-isotropic 3D region. When a 
critical value is reached, all iso-$\lambda$ manifolds cut through the origin: 
at this point, the attractor dimension undergoes a transition to the 3D 
anisotropic region where  where it scales approximately as $\text{Ha}^{-1}$. From the data, we find that for low Reynolds numbers, the exponent is $-1 \pm 0.1$, but as the Reynolds number increases the uncertainty reduced to about $-1 \pm 0.003$, which strongly suggests that the slope tends to $-1$ at large Reynolds number.\\
For higher values of Ha, another phase transition occurs 
to the 2D state, where all modes become $x-$independent. Since their associated 
eigenvalue becomes independent of Ha (from (\ref{disp})), so does the upper 
bound for the attractor dimension.\\ 
The transition from quasi-isotropic 3D to anisotropic 3D sets occurs for $d_{\rm att} \sim Ha^{3}$  in this diagram: this transition is quite gradual, and we can find the transition curve and its slope either by examining the data for the first mode with non-zero $k_x$, zero $k_y$ and $\kappa_z$ approximately 1.57, or by finding the number of modes for which $|\lambda| < \text{Ha}^2/2$. The two approaches give the results $d_{\rm att} \simeq (0.25 \pm 0.01) \text{Ha}^{3 \pm 0.003}$   and $d_{\rm att} \simeq (2.35 \pm 0.01) \text{Ha}^{3.05 \pm 0.05}$ respectively. 
 The 2D region is that of all least dissipative modes for which $k_x=0$; this time, examining the data gives a scaling of the form $d_{\rm att}\simeq (2.2 \pm 0.1) \text{Ha}^{1.03 \pm 0.03}$ for this transition.
 
A similar approach is followed, holding Ha constant and varying Re, to determine how $d_{att}$ varies with Re 
in each of these three regions. Some sample curves are plotted on Fig. \ref{attdim}. Combining both graphs, we obtain that the upper bound for the attractor dimension follows one of three scalings:\\
In the 2D regime, 
\begin{equation}
d_{att}\simeq (1.2 \pm 0.1) \text{Re}^{2.1 \pm 0.1},
\label{eq:datt_2d}
\end{equation}
In the 3D anisotropic regime,
\begin{equation}
d_{att}\simeq (1.2 \pm 0.2) \frac{\text{Re}^{4.1 \pm 0.1}}{Ha},
\label{eq:datt_3dmhd}
\end{equation}
and in the 3D quasi-isotropic regime:
\begin{equation}
d_{att}\simeq (1.7 \pm 0.5) \text{Re}^{3.1 \pm 0.1}
\label{eq:datt_3d}
\end{equation}

The scalings for $d_{\rm att}$ in 3D regimes are consistent with the upper bound 
obtained in periodic domains and with heuristic estimates for the 
number of degree of freedoms in the system as both $d_{\rm att}\sim \text{Re}^4/\text{Ha}$ in the 
limit as Re and Ha tend to infinity while remaining within the 3D regime, i.e. $Ha<<d_{\rm att}<<Ha^3$ \cite{pa03}. This result is not \emph{a priori} 
obvious from the mathematical point of view, since walls parallel to the 
magnetic fields make the spectral distribution of the modes strongly inhomogeneous, in contrast with flows in periodic domains and with channels perpendicular to the field. From the physical point of view, however, the fact that the attractor dimension is not significantly affected by the nature of 
the boundaries when Re is large enough, reflects that the number of degrees 
of freedom in the flow is mainly determined by turbulence far from the walls.
In the 2D regime, $d_{\rm att}$ understandably behaves in the same way as in the fully periodic case, since in both cases, strictly 2D modes incur no Joule 
dissipation. In channels with walls perpendicular to the magnetic field, on the other hand, the Hartmann boundary layer that develops against the wall precludes strict two-dimensionality and significant Joule dissipation occurs there so 
that $d_{\rm att}$ continues to decrease with Ha in the quasi-2D regime.\\
A remark should be made on the value of the exponent of Re in the estimates 
for the $d_{\rm att}$. In the strongly anisotropic regime, for example, heuristic
 estimate for the number of degrees of freedom of turbulence in a 
periodic box yields $d_M\sim Re^2/Ha$ \cite{pa03}, and not $d_M\sim Re^4/Ha$. 
It was previously noted that this overestimate for the exponent of Re  
 takes its roots in the loose upper bound for the inertial terms in (\ref{trdim}).
This issue is not specific to MHD flows but betrays a core difficulty 
in the derivation of tight upper bounds for attractor dimensions in 3D turbulence. Nevertheless, the exponent of Ha in the estimate for $d_{\rm att}$ coincides with the heuristic estimates in the geometries with periodic boundary 
conditions and channels perpendicular to the magnetic field in both 3D regimes. Since our 
numerical estimate shows that this exponent also remains valid in the case of 
a channel parallel to the magnetic field, it is likely to be a tight estimate 
in this case too. 
\section{Scalings for the small scales and the boundary layer thickness \label{sec:scalings}}
Expressing the evolution of a solution of the Navier-Stokes equation in terms 
of the least dissipative modes necessitates that these modes are able to resolve the smallest structures present in the flow, namely the boundary layers and 
the dissipative scales. For the solution to be faithfully represented on this basis, it must include at least the $d_{\rm att}$ least dissipative elements  
of them \cite{pdy10_jfm,kop15_jcp} (From the physical point of 
view, more dissipative modes than these are dissipated before they are able to 
transfer energy through inertia). This uniquely determines the smallest scales present in the flow $\kappa_z^{\rm max}$, $k_x^{\rm max}$ and $k_y^{\rm max}$ as well as 
the smallest and largest possible boundary layer thicknesses. Both are readily 
extracted from the ordered sequence of least dissipative modes calculated in 
Sec. \ref{sec:distrib}.

\begin{figure}[h!]
\begin{center}
\includegraphics[width=0.5\textwidth]{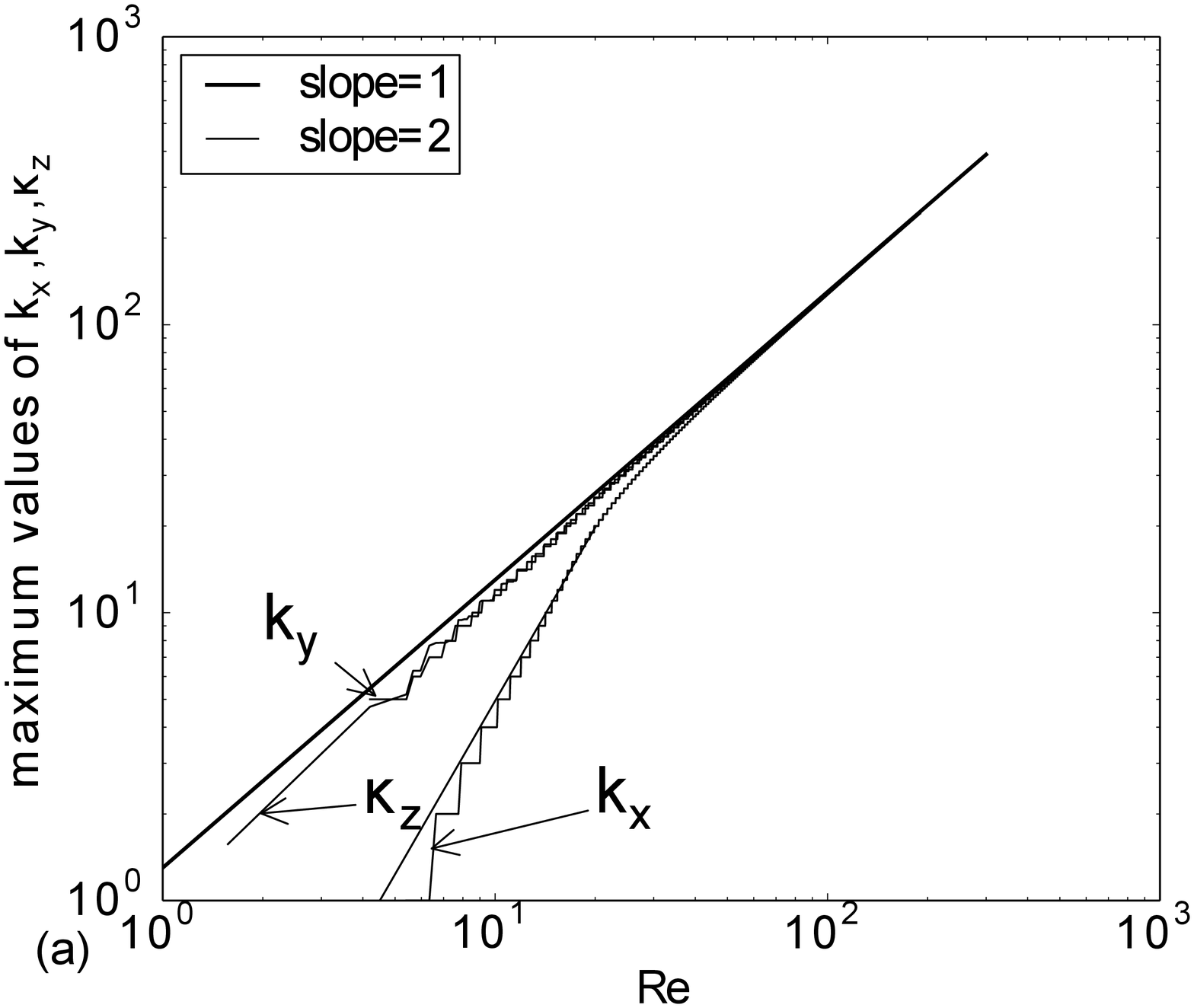}
\includegraphics[width=0.5\textwidth]{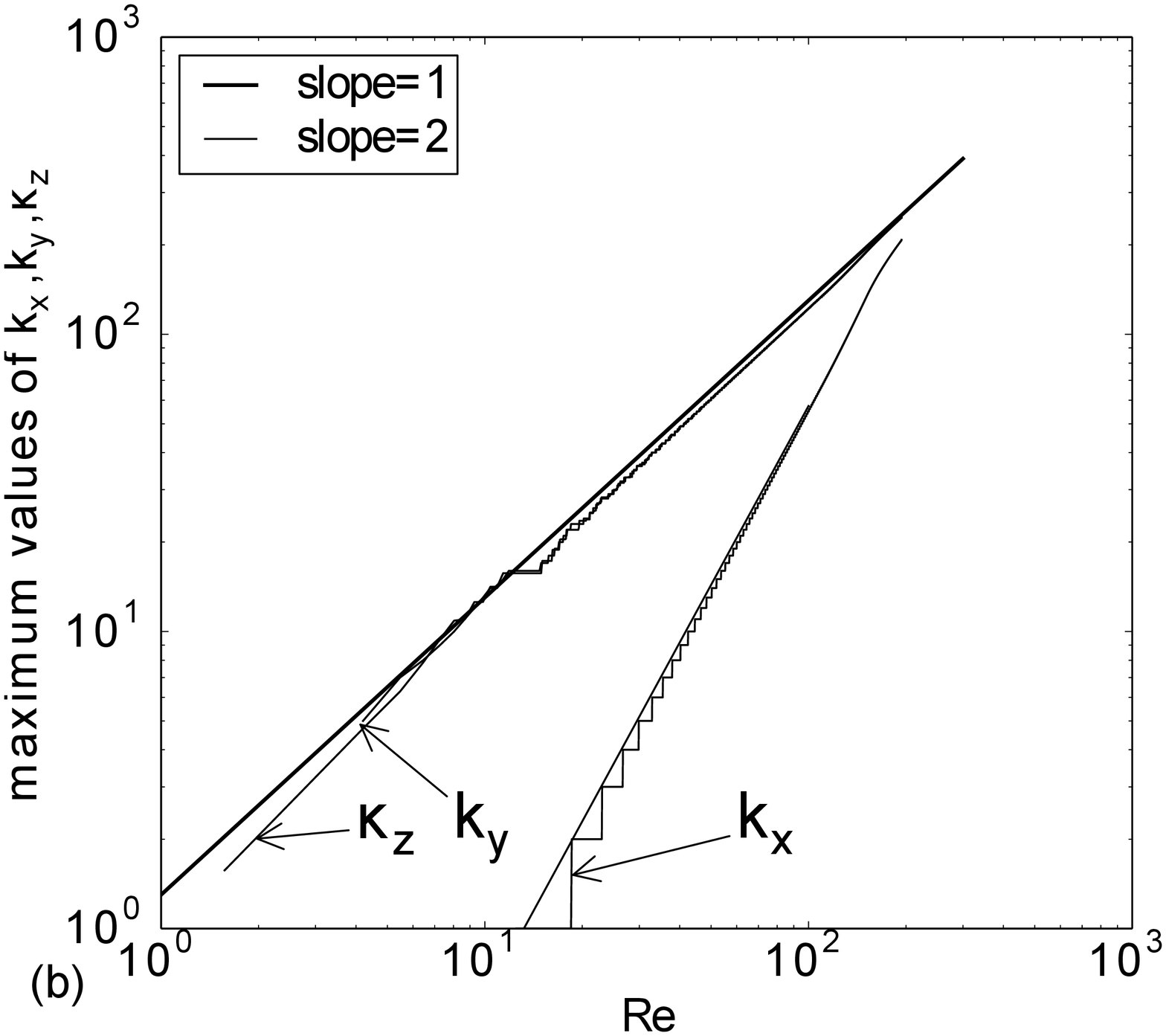}
\caption{ Maximal values of $k_x,k_y,\kappa_z$ for Ha $\approx$ 10.70 (a), and 95.40(b).}
\label{fig:kmaxvsRe1}
\end{center}
\end{figure}

\begin{figure}[h!]
\begin{center}
\includegraphics[width=0.5\textwidth]{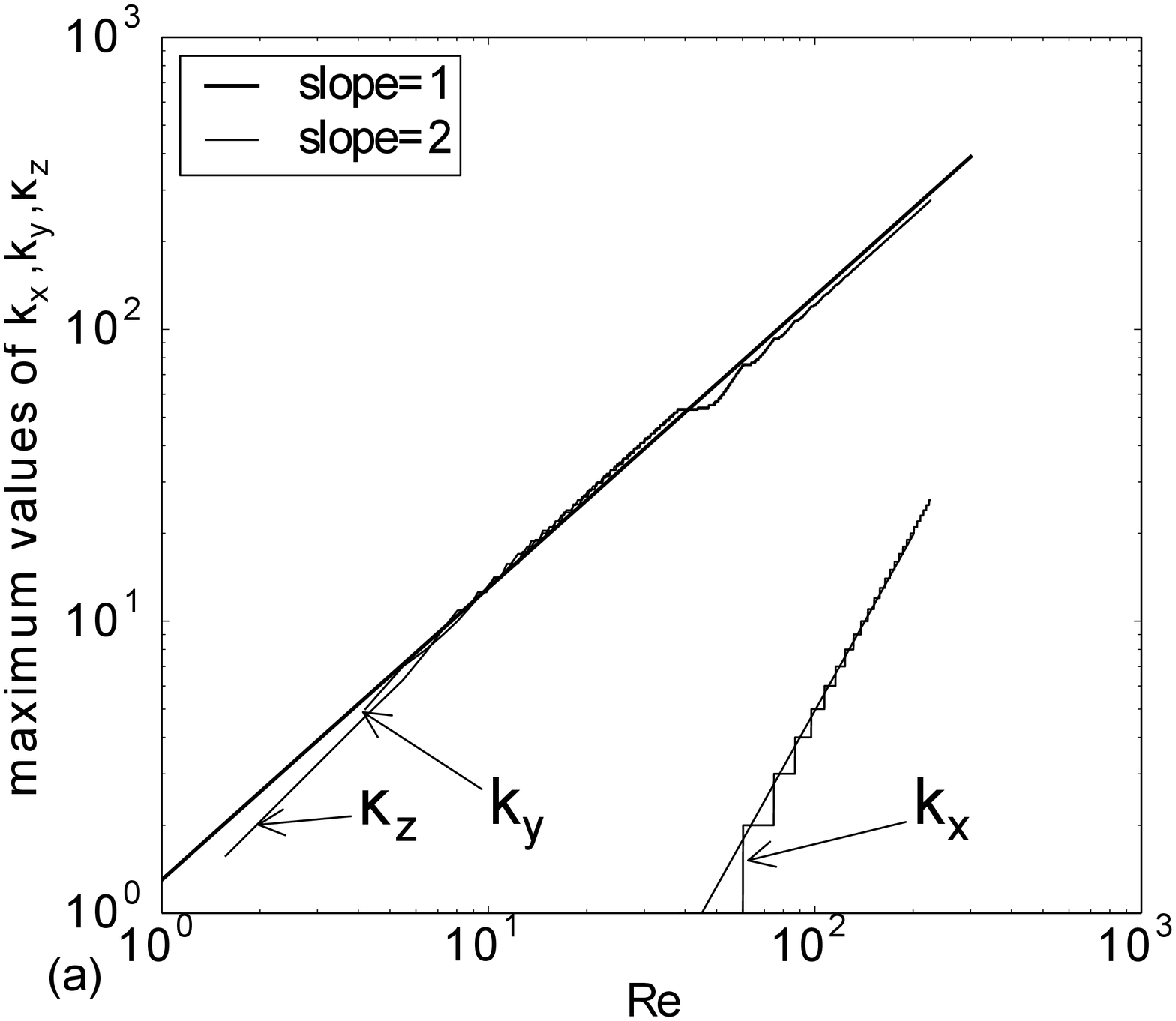}
\includegraphics[width=0.5\textwidth]{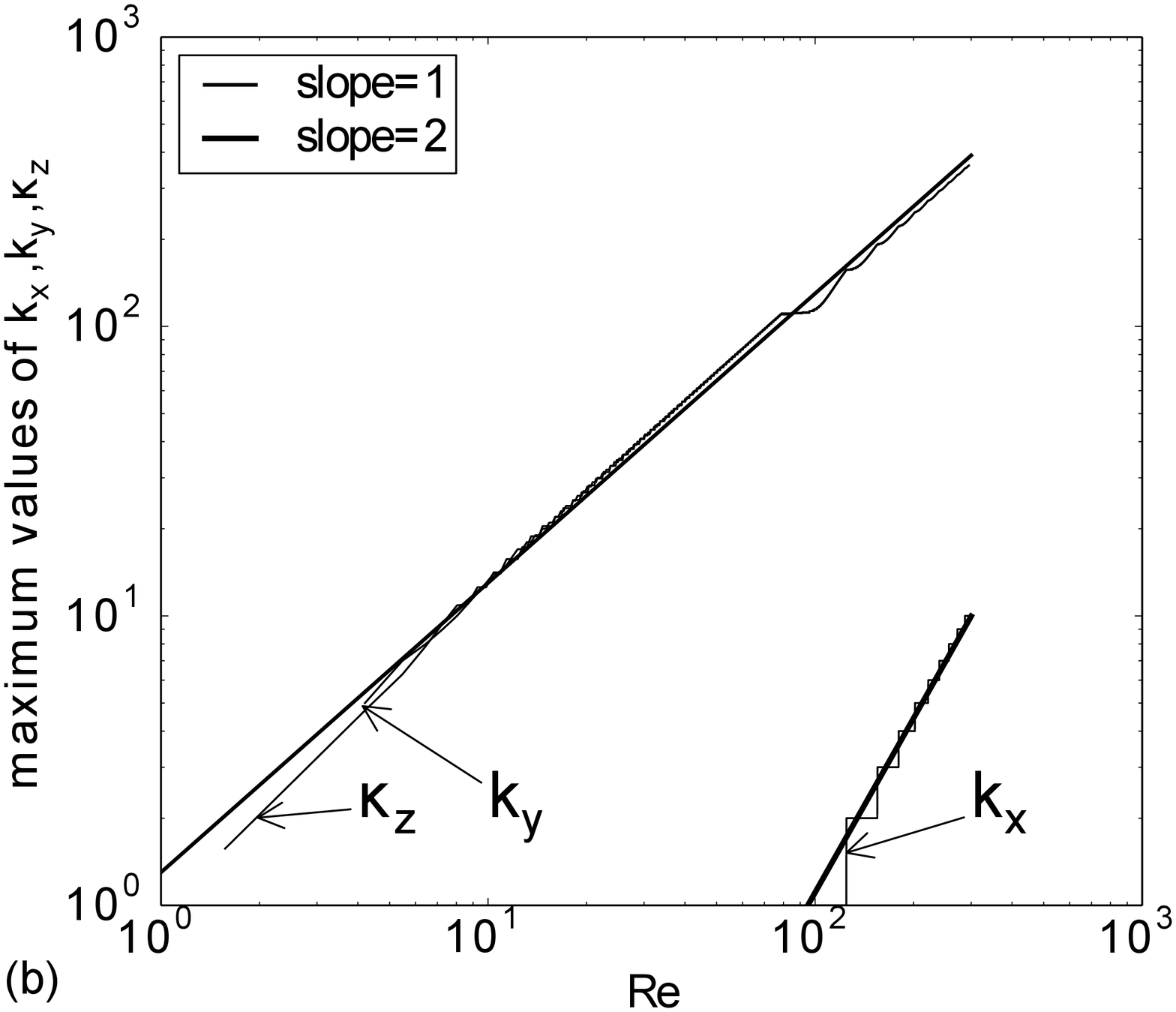}
\caption{ Maximal values of $k_x,k_y,\kappa_z$ for Ha $\approx$ 1021 (a), and 4389 (b).}
\label{fig:kmaxvsRe2}
\end{center}
\end{figure}

Let us first examine the behaviour of the small scales, shown on Figs \ref{fig:kmaxvsRe1} and \ref{fig:kmaxvsRe2}.

We see in the graphs how the maximum values of $k_x$, $k_y$ and $\kappa_z$ behave in the three regimes. For small $\text{Ha} \lesssim 1$, the system is in the quasi-isotropic 3D state for all values of Re, and we see that in this case all three of $k_x$, $k_y$ and $\kappa_z$ scale approximately as Re.
From the numerical data, the scaling is of the form 
\begin{equation}
k_x\sim k_y\sim \kappa_z\sim(1.3 \pm 0.1) \text{Re}^{1 \pm 0.05}.
\label{eq:kx_3d_m}
\end{equation}
As Ha is increased, for low values or Re we have the anisotropic 3D regime, in 
which $k_x$ is significantly less than Ha. $d_{att}$ then scales approximately as $(0.52 \pm 0.03)\text{Re}^{2 \pm 0.1}/\text{Ha}$, and as Re increases the system makes a transition to the quasi-isotropic 3D regime.The small scales are then
\begin{eqnarray}
k_x\sim(0.18 \pm 0.08) \frac{\text{Re}^{2.1 \pm 0.2}}{Ha}, \label{eq:kx_mhd_m}\\
k_y\sim \kappa_z\sim (1.3 \pm 0.1) \text{Re}^{1 \pm 0.05} \label{eq:kp_mhd_m}.
\end{eqnarray}

Finally, for Ha large enough, the fluid is in the 2D regime initially, in which $k_x$ is entirely suppressed. As Re increases, we enter the anisotropic 3D regime, with a trace of this transition appearing in the curves for $k_y$ and $\kappa_z$. For the larger values of Ha the transition to quasi-isotropic 3D takes place at too large a value of Re to be observed here.\\

\begin{figure}[h!]
\begin{center}
\includegraphics[width=0.5\textwidth]{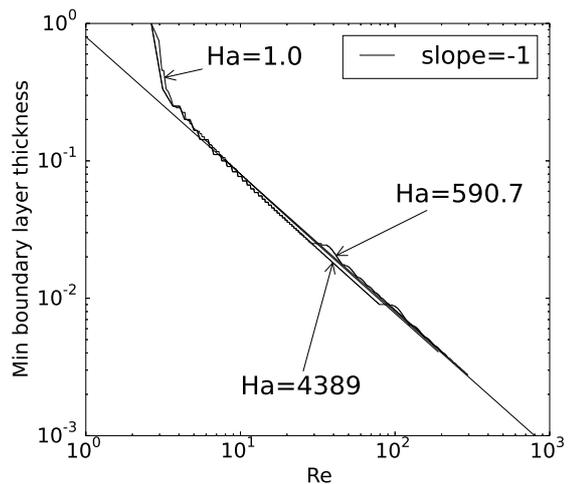}
\caption{Boundary layer thicknesses for  Ha $\approx 1.0, 590.7, 4389$.}
\label{deltavsRe}
\end{center}
\end{figure}

The boundary layers thicknesses are perhaps the most interesting because they can be expected to differ radically from the Hartmann boundary layers found in 
channels perpendicular to the magnetic field \cite{pa06_pf}: 
since the magnetic field is aligned parallel to the walls, we would not expect 
it to lead to the formation of a Hartmann layer: and indeed, the numerical 
evidence is that the minimum boundary layer thickness scales as $1/\text{Re}$, in all three regimes; here, the numerical data give a scaling law of $\delta=(0.8 \pm0.02)\text{Re}^{-1.02 \pm 0.02}$. 
The dependence on Reynolds indicates that the thinnest layer is purely viscous. 
Figure \ref{deltavsRe} shows the 
relationship for the smallest and largest Hartmann numbers considered, and one 
intermediate value. We also observe that the graph of the smallest boundary layer thickness shows a trace of the transition from 2D to anisotropic 3D flow, in the form of a discrepancy from the power law line which then settles down as Re increases, but which does not affect the asymptotic scaling. Interestingly, while the minimum boundary layer thickness does not depend on Ha, the critical value of Re at which this transition occurs, on the other hand, does.
The thickest layer, on the other hand, rapidly saturates as Re is increased. 
Unlike Hartmann layers, the layers in channels parallel to walls do not have a 
definite thickness determined by the balance between Lorentz force and viscous 
friction, even at low Re. This reflects in different modes exhibiting 
different boundary layer thicknesses at all values of Re. Since the real flow is ultimately a combination of these modes with different boundary layer thicknesses, it may not exhibit an exponential profile, unlike the Hartmann layers found in the case of the channel with transverse magnetic field. 

\section{Heuristics for the transition between turbulent regimes}
The attractor dimension represents the number of degrees of freedom of the dynamical system underlying turbulence. It can be heuristically estimated as the number of vortices  in the flow. In homogeneous hydrodynamic turbulence, Kolmogorov's law for the size of the small scales $k^{\rm max}\sim Re^{3/4}$ yields 
\begin{equation}
d_{\rm att}\sim Re^{9/4}.
\label{eq:dk41}
\end{equation}
In the anisotropic MHD regime, scalings for the small scales are usually 
obtained by assuming that anisotropy is constant along the inertial range and 
that inertial transfer is balanced by Joule dissipation at all scales in the 
inertial range, which translates into the following scaling for the anisotropy 
and the power spectral density \cite{alemany79,pa03}:
\begin{eqnarray}
\frac{k_x}{k_\perp}\sim N^{-1},\\
E(k_\perp)\sim E_0 k^{-3} \label{eq:k3spec},
\end{eqnarray} 
, where $E_0=E(k\perp=1)$, 
and since at the small scales, viscous friction becomes of the same order as 
these two effects, we obtain
\begin{eqnarray}
k_x^{\rm max}\sim \frac{Re}{Ha} \label{eq:kx_heu},\\
k_y^{\rm max}\sim\kappa_z^{\rm max}\sim k_\perp^{\rm max}\sim Re^{1/2},
\label{eq:kp_heu}\\
d_{\rm att}\sim \frac{Re^2}{Ha}. \label{eq:d_mhd_heu}
\end{eqnarray} 
The transition between the homogeneous isotropic regime and the anisotropic MHD regime, occurs when the estimates for $k_\perp$ and $k_x$ converge to the 
same value.  Whether using mathematical estimates (\ref{eq:kx_mhd_m}), (\ref{eq:kp_mhd_m}) and (\ref{eq:datt_3dmhd}), or heuristics (\ref{eq:kx_heu}), (\ref{eq:kp_heu}) and (\ref{eq:d_mhd_heu}) the number of degrees of freedom at the transition scales as $d_{\rm att}\sim Ha^3$, in line with the numerical findings of 
Sec. \ref{sec:bounds}.\\
Similarly, the transition between anisotropic MHD regime and the 2D regime 
takes place when $k_x^{\rm max}\sim 1$. Applying this condition to both the 
mathematical 
estimates (\ref{eq:datt_3dmhd}) and (\ref{eq:kx_mhd_m}), and the heuristics estimates 
(\ref{eq:d_mhd_heu}) and (\ref{eq:kx_heu}) yields the same scaling $d_{\rm att}\sim Ha$.  It is remarkable that when expressed in terms of the number of 
degrees of freedom rather than Reynolds number, the transition 
laws found from the properties of the least dissipative modes reflect 
heuristics accurately, and do not suffer from the loose estimate for the 
inertial terms. 
Nevertheless, it should be noted that 
the heuristic phenomenology for MHD turbulence discussed in this section 
is only well established 
 for values of the interaction parameter $S$ of the order of unity 
\cite{alemany79}. The authors of \cite{eckert01_ijhff} experimentally observed that the 
spectral exponent in the inertial range varied continuously but 
non-monotonously between -5/3 and -4 when $S$ spanned larger intervals from 0 
to large values.  Although the full range of these values included nearly 
isotropic regimes and quasi-2D regimes, this stresses that the $k^{-3}$ spectrum
 is not a universal feature of anisotropic MHD turbulence, unlike the $k^{-5/3}$ spectrum of isotropic, homogeneous hydrodynamic turbulence. Scalings for the small scales of MHD 
turbulence with a different spectrum are however not known. Equally, the scaling $d_{\rm att}\sim Ha^{-1}$ is asymptotic and Fig. \ref{attdim} shows that it may be imperfectly verified away from the middle range of the anisotropic regime.\\
Finally, it should be noticed that the phenomenology discussed in this section 
applies regardless of the boundary conditions, and therefore to Hartmann flows 
and flows in 3D periodic domains \cite{pa03,pa06_pf}. The most remarkable 
aspect about the case of a channel flow with a spanwise magnetic field is that 
the same phenomenology applies to it despite a very different spectral 
distribution of eigenmodes. Although not surprising from the physical point of 
view, this property is anything but straightforward from the mathematical point 
of view.
\section{Concluding remarks}
The sequence of least dissipative modes for a channel flow in a homogeneous magnetic field parallel to the walls has been derived. This achieves the first step 
towards spectral DNS of MHD flows in this configuration based on this functional basis. This promising method was shown to partially lift the cost of meshing the very thin boundary layers in MHD channel flows perpendicular to the externally applied magnetic field. In the present case, Hartmann 
layers are not present. This implies that when the flow is 3D but strongly 
anisotropic, the number of modes required to 
represent the flow completely can be expected to come close to the actual 
attractor dimension $d_{\rm att}$, as in the case of periodic flows \cite{pdy10_jfm}. An upper bound for it was found to scale as 
$\text{Re}^4/\text{Ha}$, but $d_{\rm att}$ itself is heuristically expected to scale as 
$\text{Re}^2/\text{Ha}$, suggesting that the upper bound we find is tight as far as the exponent of Ha is concerned but not that of Re. Either way, $d_{\rm att}$ significantly decreases with Ha and so using the 
least dissipative modes in spectral DNS should incur 
significant computational savings.\\  
The success of such a numerical approach relies on the ability of these modes 
to faithfully represent the physical properties of the flow. In this respect 
the least dissipative modes have been shown to recover most of the known 
attributes of MHD turbulence in a channel parallel to the magnetic field:

\begin{itemize}[leftmargin=0.2cm]
\item In regimes where the flow is 3D, turbulence far from the 
wall determines the number of degrees of freedom of the flow. The ensuing 
scalings for the attractor dimension, small scales along and across the magnetic fields, and Joule cone half-angle are essentially the same as those for turbulence in a periodic domain. These are all finely recovered by the set of least dissipative modes. In the most important case of 3D anisotropic flow, these scalings for the small scales were:

\begin{displaymath}
\kappa_z\simeq k_y \simeq 01.3 \text{Re} \qquad k_x\simeq\frac{\text{Re}^{2}}{2 \text{Ha}} 
\end{displaymath}
\item The modes spread into two families: Orr-Sommerfeld modes, which have a velocity component across the channel and Squire modes which do not.

\item The spectral distribution of the least dissipative modes is strongly inhomogeneous, because of the presence of pairs of OS modes with imaginary eigenvalues. This effect is due to the presence of walls parallel to the magnetic field but does not affect the main scalings for the attractor dimension: this important result is \emph{a priori} far from obvious from the mathematical 
point of view but reflects that high-Re turbulence is not strongly affected by the walls in the present geometry (see above).
\item The maximum and minimum thicknesses of the boundary layers associated to the least dissipative modes along the walls are essentially 
independent of the external magnetic field and depend on Re only, as one would expect for a magnetic field parallel to the walls.
\end{itemize}
The authors  gratefully acknowledge financial support from the Leverhulme Trust (Grant Ref. F00/732J).

\appendix
\section{Sporadic resonant modes \label{sec:resonant}}
Only the cases considered above provide modes for arbitrarily chosen Hartmann number. Of the several other possible cases (both roots of the quadratic equation in $K^2$ real, one or both 0, both roots of the same magnitude) some have no modes, others have only modes in which the Hartmann number is precisely determined by $k_x$ and $k_y$ and the values of $K_1$, $K_2$. We tabulate the possibilities as follows:
\begin{enumerate}
\item$K_1=0$, $K_2^2=-\mu^2$: $\mu=n \in \mathbb{Z}$, and $2\text{Ha}^2=(k^2+n^2)/k_x k^2$
\item $K_1^2=-K_2^2=\mu^2$, where $\mu \in \mathbb{R}$: $\mu$ must satisfy the equation $\tan(\mu)=\pm\tanh(\mu)$, and $2\text{Ha}^2=(k^2-\mu^2)/k_x k^2$.
\item All other cases: no nontrivial modes.
\end{enumerate}

\section{Eigenbasis of the Dissipation operator \label{sec:basis}}

To begin with, we see the various OS modes, which themselves split up into several cases.

First, we have the case where the roots of (\ref{eq:charact}) are $\pm 1/\delta$ and $\pm i \kappa_z$, where $1/\delta \neq \pm \kappa_z$, and $k_x$ and $k_y$ are not both zero.

If $1/\delta \tanh 1/\delta = -\kappa_z \tan \kappa_z$ then $Z_z(z)$ is given by
\[
Z_z(z)=-\cos(\kappa_z)\cosh(z/\delta)+\cosh(1/\delta)\cos(\kappa_z z)
\]

If neither $k_x$ nor $k_y$ is $0$, then $Z_x(z)$ and $Z_y(z)$ are given by
\[
\begin{split}
Z_x(z)&=i\frac{k_x ( \kappa_z^3 \cosh(1/\delta) \sin \kappa_z-1/\delta^3 \cos \kappa_z \sinh(1/\delta)) }{(1/\delta^2+\kappa_z^2)(k_x^2+k_y^2)}\\
&\times \left(\frac{\sinh(z/\delta)}{\sinh(1/\delta)}- \frac{\sin(\kappa_z z)}{\sin \kappa_z}\right)\\
Z_y(z)&=\frac{1}{k_y}(iZ'_z(z)-k_xZ_x(z))
\end{split}
\]

If $1/\delta \tan \kappa_z = \kappa_z \tanh 1/\delta$, then $Z_z(z)$ is given by
\[
Z_z(z)=-\sin\kappa_z \sinh(\kappa_z z) +\sinh(1/\delta) \sin(\kappa_z z)
\]

If neither $k_x$ nor $k_z$  is $0$, then $Z_x(z)$ and $Z_y(z)$ are given by
\[
\begin{split}
Z_x(z)&=i\frac{k_x(1/\delta^3 \sin \kappa_z \cosh(1/\delta)  + \kappa_z^3 \sinh (1/\delta) \cos \kappa_z)}{(k_x^2+k_y^2)(1/\delta^2+\kappa_z^2)}\\
&\times \left( \frac{\cos(\kappa_z z)}{\cos \kappa_z} - \frac{\cosh(z/\delta)}{\cosh(1/\delta)} \right)\\
Z_y(z)&=\frac{1}{k_y}(iZ'_z(z)-k_xZ_x(z))
\end{split}
\]
If $k_x=0$,
\[
Z_x(z)=0, \qquad Z_y(z)=iZ'(z)/k_y
\]
and if $k_y=0$,
\[
Z_y(z)=0, \qquad Z_x(z)=iZ'(z)/k_x
\]

In each of these cases the eigenvalue is given by
\[
\lambda=\frac{1}{2}\left(\frac{1}{\delta^2} - \kappa_z^2 \right) - k^2.
\]

Next, we have the case where the roots of (\ref{eq:charact}) are $\pm i\kappa_z$ and $\pm i \tilde{\kappa}_z$, where
$\kappa_z \neq \pm \tilde{\kappa_z}$, and $k_x$ and $k_y$ are not both zero.

If 
\[
\tilde{\kappa}_z \tan \tilde{\kappa}_z = \kappa_z \tan \kappa_z
\]
then $Z_z(z)$ is given by
\[Z_z(z)=-\cos \kappa_z \cos(\tilde{\kappa}_z z)+\cos \tilde{\kappa}_z \cos (\kappa_z z)
\]
If neither $k_x$ nor $k_y$ are zero, then $Z_x(z)$ and $Z_y(z)$ are given by
\[
\begin{split}
Z_x(z&)=i\frac{k_x(\tilde{\kappa}_z^3 \cos \kappa_z \sin \tilde{\kappa}_z  - \kappa_z^3 \cos \tilde{\kappa}_z \sin \kappa_z )}{(k_x^2+k_y^2)(\tilde{\kappa}_z^2-\kappa_z^2)} \\ 
&\times \left(\frac{\sin(\tilde{\kappa}_z z)}{\sin \tilde{\kappa}_z} - \frac{\sin(\kappa_z z)}{\sin \kappa_z}\right)\\
Z_y(z)&=\frac{1}{k_y}(iZ'_z(z)-k_xZ_x(z))
\end{split}
\]

If
\[
\tilde{\kappa}_z \tan \kappa_z =\kappa_z\tan \tilde{\kappa}_z
\]
then $Z_z(z)$ is given by
\[
Z_z(z)=-\sin(\kappa_z)\sin(\tilde{\kappa}_z z)+\sin(\tilde{\kappa}_z) \sin(\kappa_z z)
\]
and if neither of $k_x$ nor $k_y$ is zero then $Z_x(z)$ and $Z_y(z)$ are given by
\[
\begin{split}Z
_x(z)&=i\frac{k_x(\kappa_z^3 \sin \tilde{\kappa}_z \cos \kappa_z-\tilde{\kappa}_z^3 \sin \kappa_z \cos \tilde{\kappa}_z)}{(k_x^2+k_y^2)(\tilde{\kappa}_z^2-\kappa_z^2)}\\
&\times \left(\frac{\cos(\tilde{\kappa}_z z)}{\cos \tilde{\kappa}_z}- \frac{\cos(\kappa_z z)}{\cos \kappa_z}\right)\\
Z_y(z)&=\frac{1}{k_y}(iZ'_z(z)-k_xZ_x(z))
\end{split}
\]
Just as before, if $k_x=0$,
\[
Z_x(z)=0, \qquad Z_y(z)=iZ'(z)/k_y
\]
and if $k_y=0$,
\[
Z_y(z)=0, \qquad Z_x(z)=iZ'(z)/k_x
\]
In each of these cases the eigenvalue is given by
\[
\lambda = -\frac{1}{2} (\kappa_z^2+\tilde{\kappa_z}^2)-k^2.
\]

Finally, we have the Squire modes: for these modes, $k_x=k_y=0$, and we have $Z_z(z)=0$, and for each positive integer $n$ there are modes
\[
Z_{x,y}= \cos((n+1/2)\pi z)
\]
with $\lambda = -\frac{1}{2}(n+1/2)^2\pi^2$ and
\[
Z_{x,y}(z)=\sin(n \pi z)
\]
with $\lambda=-\frac{1}{2}n^2\pi^2$.

\bibliographystyle{apsrev4-1}
\bibliography{fullbiblio}

\end{document}